\documentclass[twocolumn]{aastex63}

\usepackage{graphics,epsf}
\usepackage{amsmath}                
\usepackage{amsfonts}               
\usepackage{amssymb}                
\usepackage{epsfig}                 
\usepackage{float}
\usepackage{color}
\usepackage{tabularx}
\usepackage{comment}
\usepackage{makecell}

\def \cm{~\rm{cm}}
\def \s{~\rm{s}}
\def \km{~\rm{km}}

\def \K{~\rm{K}}

\def \g{~\rm{g}}
\def \G{~\rm{G}}

\def \erg{~\rm{erg}}

\def \kpc{~\rm{kpc}}

\usepackage{xcolor}
\definecolor{redak}{rgb}{0.9,0.15,0.05}

\begin{document}
\title{On the nature of the planet-powered transient event ZTF SLRN-2020}

\author[0000-0003-0375-8987]{Noam Soker}
\affiliation{Department of Physics, Technion, Haifa, 3200003, Israel; soker@physics.technion.ac.il}



\begin{abstract}
The Red Nova ZTF SLRN-2020 is the third transient event with properties that are compatible with the merger of a planet with a main sequence (or close to) star on a dynamical timescale. While the two first transient events occurred in young stellar objects, ZTF SLRN-2020 occurred in an old system. Nonetheless, I show that the three star-planet intermediate luminosity optical transients (ILOTs, also termed Red Novae) occupy the same area in the energy-time diagram of ILOTs. Based on models for ILOTs that are power by stellar binary interaction I suggest that the planet in ZTF SLRN-2020 launched jets at about its escape speed before it was engulfed by the star. Interestingly, the escape speed from the planet is similar to the orbital speed of the planet. This leads to an outflow with a very low terminal velocity, much below the escape velocity from the star, and in concentration around $\approx 45 ^\circ$ to the equatorial plane. As well, the planet might have lost back some of the accreted mass just before engulfment, forming an accretion disk around the star. This disk might have launched jets during the main outburst of the event. The jets form a bipolar expanding nebula.   
\end{abstract}


\section{Introduction}
\label{sec:Introduction}

The recently analyzed \citep{Deetal2023} interesting transient event ZTF SLRN-2020 is the fourth intermediate luminosity optical transient (ILOT; or Red Nova) that was claimed to be powered by star-planet interaction.\footnote{I refer to all gravitational-powered transients as ILOTs, not including supernova nor dwarf novae (e.g., \citealt{Berger2009, KashiSoker2016, MuthukrishnaetalM2019}). Other researchers use other terms (e.g., \citealt{Jencsonetal2019, PastorelloMasonetal2019, PastorelloFraser2019}. The ILOT class includes in it the subclass of Red Novae and some similar sub-classes. }  

The first claim by \cite{RetterMarom2003} and \cite{Retteretal2006} that V838Mon was powered by star-planet interaction was refuted on the ground of energy considerations (e.g., \citealt{SokerTylenda2003}). 

The second claim for an ILOT powered by a star-planet interaction is that the unusual outburst of the young stellar object (YSO) ASASSN-15qi was powered by the accretion of a tidally-disrupted Jupiter-like planet \citep{KashiSoker2017Planet}. The accretion via an accretion disk powered the event, probably by launching jets. \cite{Kashietal2019Galax} made the third claim and argued that the $\approx 800$ days-long eruption of the young stellar object ASASSN-13db was powered by engulfment of the remains of a planet that was tidally-shredded by the YSO. These last two claims are not refuted (yet), and so the new claim by \cite{Deetal2023} is the third claim for a star-planet-powered Red Nova that still holds. 

From their analyses of the observations \cite{Deetal2023} deduce that a star of a mass $M_1\simeq 0.8-1.5 M_\odot$ on the main sequence or early sub-giant phase (Hertzsprung gap) and with a radius of $R_1\simeq 1-4 R_\odot$ engulfed a planet to power this event that radiated in the first 150 days an energy of $E_{\rm rad} \simeq 6.5 \times 10^{41} (d/4 \kpc)^2 \erg$, where $d$ is the distance to this ILOT.  
\cite{Deetal2023} argue for a planet mass of $M_{\rm p} \simeq 0.01 M_\odot$. They further deduce from the infrared that the interaction started at least $\approx 7$~months before the main outburst, and that the interaction expelled dust with a velocity of $v_{\rm d} \simeq 35 \km \s^{-1}$. 

In this short study I further analyze the new event ZTF SLRN-2020. I place it on the energy-time diagram of ILOTs and compare it with predictions of previous studies and discuss the common properties and differences from the two earlier ILOTs powered by star-planet interaction (section \ref{sec:Comparing}). 
I then suggest (section \ref{sec:Jets}) that the slowly expanding dust was launched from an accretion disk around the planet. I do not study other theoretical aspects that were worked out by earlier studies (e.g., \citealt{Bearetal2011, Metzgeretal2012, Yamazakietal2017, Gurevichetal2022,OConnoretal2023}). I summarize in section \ref{sec:Summary}.  

\section{Comparing ZTF SLRN-2020 with other star-planet ILOTs}
\label{sec:Comparing}

I consider the duration of the star-planet ILOTs and their total energy that includes the radiated energy and the kinetic energy of the ejecta. I end this section with the discussion of the main difference between them. 

\begin{figure*}
\includegraphics[trim= 0.0cm 0.0cm 0.0cm 0.0cm,clip=true,width=1.0\textwidth]{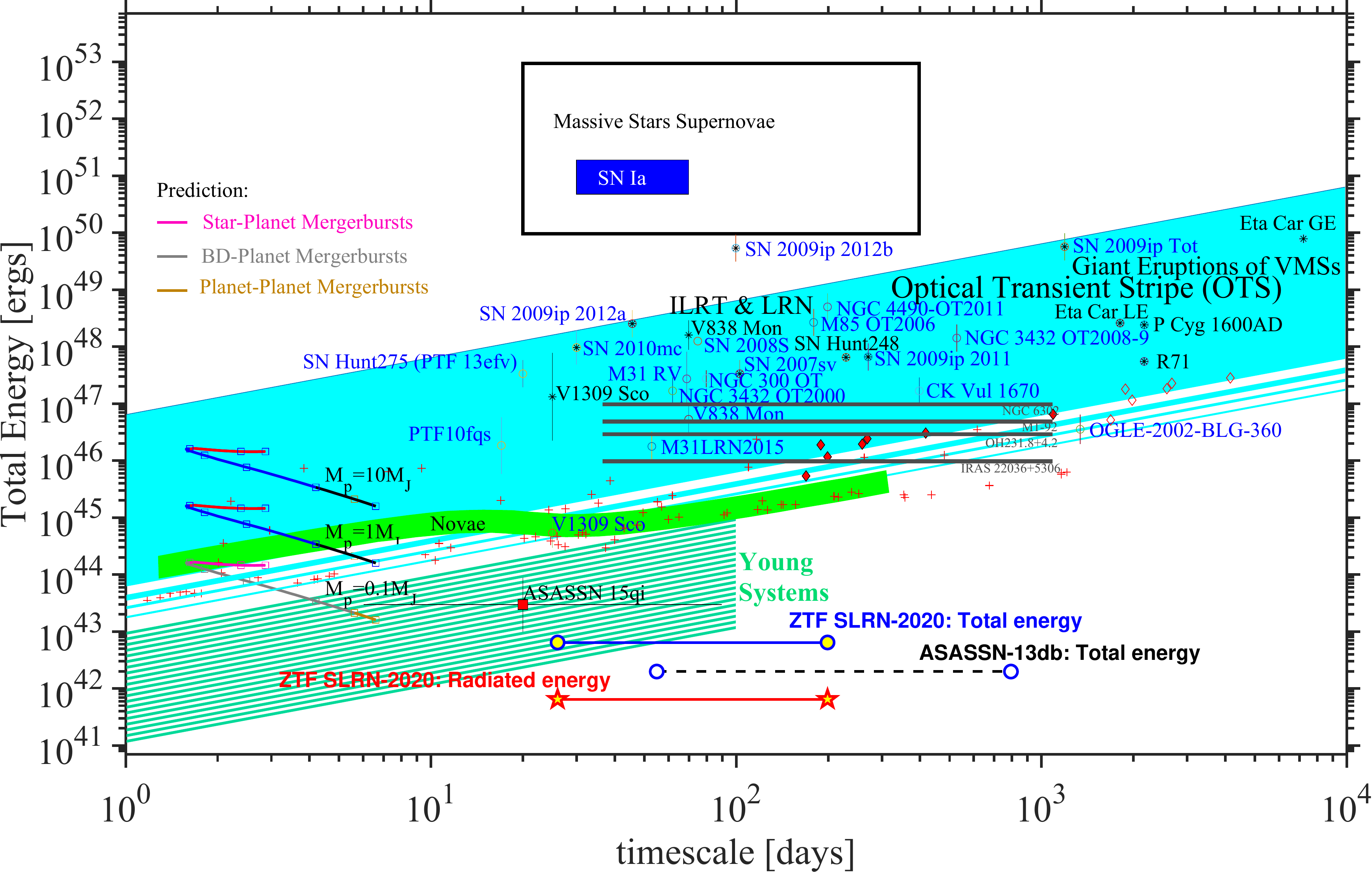}
\vskip +0.00 cm
\caption{Observed transient events on the energy time diagram adapted from \cite{KashiSoker2017Planet} (where more details can be found). Blue empty circles represent the total (radiated plus kinetic) energy
of the observed transients as a function of the duration of their eruptions, i.e., usually the time for the visible luminosity to decrease by 3
magnitudes. The three lines on the left are theoretical estimates for ILOTs powered by planet-brown dwarf (BD) interaction \citep{Bearetal2011}. 
\cite{KashiSoker2017Planet} place the ILOT ASASSN-15qi (observational data from \citealt{Herczegetal2016}) and argued it was powered by a star-planet interaction. I added ASASSN-13db (observational data from \citealt{SiciliaAguilaretal2017}) with its estimated total energy that \cite{Kashietal2019Galax} claimed to have been powered by a star engulfing planet debris, and ZTF SLRN-2020 from \cite{Deetal2023} who argued it was powered by a star engulfing a planet. Both the radiated energy (yellow-filled red stars) and the crudely estimated total energy (radiation + kinetic; two yellow-filled blue circles) are shown with the timescale range of ZTF SLRN-2020 (see text). The timescale range is not due to uncertain observations, but rather in the way that one defines the timescale, total duration or decline rate. The lower-left part (hatched in green) is the extension that \cite{KashiSoker2017Planet} made to include YSO-planet systems. The new observations suggest that systems of a well evolved main sequence star and a planet also populate this region. Abbreviation: BD: brown dwarf; GE: Great Eruption; ILRT: intermediate luminosity red transient; LRT: luminous red transient; SN: supernova; VMSs: very massive stars;  
}
\label{fig:ETD}
\end{figure*}

The total radiated energy of ZTF SLRN-2020 is $E_{\rm rad} \simeq 6.5 \times 10^{41} (d/4 \kpc)^2 \erg$, and the decline time by three magnitudes is about 200~days \citep{Deetal2023}. \cite{Deetal2023} consider two limits for the duration of the event. Its lightcurve plateau duration, $\simeq 26$~days, and the time it radiated 90\% of its the total radiated energy, $103 \pm 20$~days. The total radiate energy divided by the luminosity during the plateau $L_{\rm p} \simeq 1.1 \times 10^{35} (d/4 \kpc)^2 \erg \s^{-1}$ \citep{Deetal2023} gives a timescale of $\simeq 70$~days. I take the range to include all these timescales, namely, from $26$~days to $200$~days. I mark these two end with yellow-red stars on Fig. \ref{fig:ETD}. Fig. \ref{fig:ETD} that is adapted from \cite{KashiSoker2017Planet} presents many ILOTs in a plane of their total energy versus their typical timescale. Note that the energy is the total energy, including radiation and kinetic energy.  

When there is no information on the kinetic energy the structure of this diagram assumes that the total energy is ten times the radiated energy. This estimate gives here a total energy of $E \simeq 6.5 \times 10^{42}$ (for a distance of $d=4 \kpc$) for ZTF SLRN-2020. 
\cite{Deetal2023} crudely estimate an ejecta mass and velocity $M_{\rm ej} \approx 3 \times 10^{-5} M_\odot$ and $v_{\rm ej} \approx 100 \km \s^{-1}$. This gives a kinetic energy of $E_{\rm ej} \approx 3 \times 10^{42} \erg$. The estimated total energy from observations is therefore $E_{\rm ej} + E_{\rm rad} \approx 3.6 \times 10^{42} \erg$. Therefore, the value of $E \simeq 6.5 \times 10^{42}$ that I take here is a reasonable upper value of the total energy (and there is the additional uncertainty of the distance). I mark the location of  ZTF SLRN-2020 on Fig. \ref{fig:ETD} with $E \simeq 6.5 \times 10^{42}$  and the range of timescales by the two yellow-filled blue circles and a line connecting them.   

\cite{Kashietal2019Galax} analyzed the ILOT ASASSN-13db that was observed by \cite{SiciliaAguilaretal2017}. The total duration of high emission lasted $\approx 800$~day, ending with a decline that lasted $\approx 55$~days. The total radiated energy during the (more or less) plateau of $\approx 800$~days is $E_{\rm rad} \approx  2 \times 10^{41} \erg$. The total energy of the event is about an order of magnitude larger.  \cite{Kashietal2019Galax} scale it with $\approx 10^{42} \erg$. I take it here to be ten times lager at $E \simeq 2 \times 10^{42}$. I mark this energy with the span of the timescale of 55 to 800 days. This timescale range is not the uncertainty in observations, but rather the unclear way by which the event timescale should be defined, i.e., should we take only the decline phase by three magnitudes or so, or the entire duration, or the plateau duration, etc. 

From the locations of the three ILOTs (Red Novae) claimed to be powered by star-planet interaction, ASASSN-15qi, ASASSN-13db, and ZTF SLRN-2020, on the Energy-Time diagram we can learn the following. (1) V838 Mon is much above their location, and it was not powered by a star-planet interaction, contrary to the claim by \cite{RetterMarom2003}. (2) The three star-planet ILOTs occupy a relatively well defined area in the energy-time diagram that is clearly below those that are thought to be powered by stellar-binary systems.
(3) The star-planet ILOTs have typical timescales that are much longer than the dynamical time of the star, by about two orders of magnitude and more. This might suggest a powering process that is much longer than the dynamical time scale. An accretion disk that is formed by the planet might have this property (e.g., \citealt{Bearetal2011}). 
(4) The timescale of the star-planet ILOTs is not necessarily well defined, as a plateau phase in their lightcurve might be very long. Nonetheless, their decline phase might have a similar behavior as regular ILOTs (e.g., \citealt{Kashietal2019Galax}). 
(5) The recently added star-planet ILOT, 
ZTF SLRN-202 \citep{Deetal2023}, does not seem to be a YSO (or even a pre-main sequence star) as the other two are. Despite that it occupies the same general area of the other two star-planet ILOTs and close to the area that \cite{KashiSoker2017Planet} mark to be due to star-planet interaction in planetary systems of YSOs (green lines on the lower left of the diagram). This suggests that there are common powering processes to these three ILOTs. A common properties might results from accretion disks that launch jets. I turn to study this possibility in section \ref{sec:Jets}. 

There are some differences between the three ILOTS, in particular the multi-outbursts that ASASSN-15qi and ASASSN-13db had, but not ZTF SLRN-2020. ASASSN-13db experienced a shorter and fainter outburst in 2013 \citep{Holoienetal2014}, about a year before the beginning of the main outburst \citep{SiciliaAguilaretal2017}. \cite{Kashietal2019Galax} attribute this early outburst to a typical outburst of YSOs. The ILOT ASASSN-15qi experienced an outburst in 1976, 39 years prior to the main outburst \citep{Herczegetal2016}. There is not enough data to classify the early outburst  either as an ILOT with another planet or as a typical YSO outburst. In any case, it seems that the two early outbursts in the YSO ILOTs, which ZTF SLRN-2020 was not observed to have, result from ASASSN-15qi and ASASSN-13db being YSOs.

\section{The roles of jets}
\label{sec:Jets}

Observations and theoretical arguments suggest that jets play major roles in many ILOTs (Red Novae). 
Observations of ILOTs that have a spatially resolved ejecta show the ILOTs to possess bipolar structures that strongly hint at shaping by jets. 
These include the Great Eruption of Eta Carinae \citep{DavidsonHumphreys1997}, a luminous blue variable (LBV) with its bipolar Homunculus nebula, V4332~Sgr that has a bipolar structure \citep{Kaminskietal2018}, 
and Nova~1670 (CK~Vulpeculae) with a 350-years old bipolar nebula \citep{Sharaetal1985} that has an S-morphology \citep{Kaminskietal2020Nova1670, Kaminskietal2021Nova1670}, a morphological type that must be shaped by jets.  In a new study, \cite{Mobeenetal2023} report that the ejecta (nebula) of V838 Mon is bipolar. 

Theoretically, jets are very efficient in powering ILOTs and more efficient than equatorial ejecta (\citealt{Soker2020JetsILOT}; for powering by equatorial ejecta see, e.g., \citealt{Pejchaetal2016a, Pejchaetal2016b, MetzgerPejcha2017}). Specific studies show that jets can account for the lightcurves of at least some ILOTs (e.g., \citealt{Soker2020JetsILOT, SokerKaplan2021}). These studies were aiming at ILOTs powered by stellar binary interaction. I turn to suggest that the planet in ZTF SLRN-2020 also launched jets (or disk-wind).   

I consider the following parameters that \cite{Deetal2023} infer for the planetary system ZTF SLRN-2020. A stellar mass of $M_1 \simeq 1M_\odot$, a stellar radius of $R_1\simeq 1-4 M_\odot$, and a planet mass of $M_{\rm p} \simeq 0.01M_\odot$.
For the planet to expel mass from the system or to accrete mass it should orbit very close to the stellar surface. For a solar mass and an orbit at $a= 2 R_\odot$ the orbital velocity is $v_{\rm orb}=310 \km \s^{-1}$. The radius of the planet of the above mass is $R_{\rm p} \simeq 0.1 R_\odot$ (e.g., \citealt{Bashietal2017}), and so the escape velocity from the planet is $v_{\rm p,es} \simeq 200 \km \s^{-1}$. 

The ILOT ZTF SLRN-2020 had months of pre-outburst activity, including mass loss \citep{Deetal2023}. I consider a process by which the planet accreted mass and launched jets (or a bipolar disk wind) that powered the pre-outburst activity. As is the case with stellar winds, the typical terminal velocity of jets is about the escape velocity from the object that launches the jets (e.g., \citealt{Livio2009}). The properties of jets might change over short timescales relative to their total activity time period.  Some parts in the jets during some jet-launching episodes might have terminal velocities that are larger than the escape velocity, i.e., $v_{\rm jet, m} = \beta v_{\rm p,es}$ with $\beta>1$. For the parameters I take above, the escape velocity from the star at the location of planet is  $v_{\rm 1,es} = 2^{1/2} v_{\rm orb} \simeq 440 \km \s^{-1}$. Because the jets are launched perpendicular to the orbital plane, the terminal velocity of the fastest parts of the jets relative to the star is
\begin{equation}
v_{\rm jet,m,1} = \sqrt{(\beta v_{\rm p, es})^2 + v^2_{\rm orb}}.
\label{eq:JetMaxV}
\end{equation}
These segments of the jets escapes the system if 
$v_{\rm jet,m,1} > v_{\rm 1,es} $, which reads $\beta > v_{\rm orb} / v_{\rm p,es} \simeq 1.5$. The property of this system that the escape velocity from the planet is of the same order of magnitude as the escape velocity from the system implies that some jets segments can reach the escape velocity, but not by much. These jet segments barely escape the star. If they do, their final outflow velocity from the system is much smaller than the escape velocity from the system. 

I suggest that this pre-outburst jet activity 
explains the slow outflow velocity of the dust $\approx 35 \km \s^{-1} \ll v_{\rm orb}$ \citep{Deetal2023}. Because the outflow velocity from the planet of the escaping gas/dust, which is perpendicular to the orbital plane, is about equal to the orbital velocity, the outflow direction of the escaping dust/gas in this case is about $\theta_{\rm j}  \approx 45 ^\circ$ to the equatorial plane. This forms a bipolar outflow morphology, in addition to possible concentration of outflowing gas/dues in the equatorial plane. 

The accreted mass onto the planet forms an outer layer around the planet. When the planet spirals-in closer to the stellar surface, the planet might lose this material back to the star, hence forming an accretion disk around the star. This accretion disk might launch much faster jets. These jets might play a role in powering the main outburst, as was suggested for ILOTs power by stellar binary systems. 

If we take the lowest mass range of masses that \cite{Deetal2023} discuss for the planet in ZTF SLRN-2020, $M_{\rm p} \simeq 10^{-4} M_\odot$, then we are in a different regime. The average density of such planets is $< 1 \g \cm^{-1}$ and they might be tidally disrupted by a sun like star on the main sequence (e.g., \citealt{Yamazakietal2017}). This brings the scenario to be much more similar to what \cite{KashiSoker2017Planet} suggested for the YSO-planet system ASASSN-15qi and \cite{Kashietal2019Galax} suggested for the YSO-planet system ASASSN-13db. However, \cite{Deetal2023} consider the tidal disruption of the planet to be unlikely.  
      
\section{Summary}
\label{sec:Summary}

The goal of this study is to group the newly analyzed \citep{Deetal2023} star-planet ILOT (Red Nova) ZTF SLRN-2020 with the two other star-planet ILOTs, 
ASASSN-15qi \citep{KashiSoker2017Planet} and ASASSN-13db \citep{Kashietal2019Galax}. 
I added ASASSN-13db and ZTF SLRN-2020 to the energy-time diagram of ILOTs (Fig. \ref{fig:ETD}). These three star-planet ILOTs occupy a well defined area in that diagram, below all other ILOTs and below classical novae. They are in the general area that was marked as YSO ILOTs in young star-planet interactions. The new event ZTF SLRN-2020 is not a YSO-planet system, but nonetheless located in the same area. This might points to a similar powering mechanism. 

I suggested (section \ref{sec:Jets}) that one of the common ingredients might be the launching of jets by the star. For the ILOT ASASSN-15qi that occurred in a young planetary system \cite{KashiSoker2017Planet} suggested that the star tidally disrupted the planet, forming an accretion disk that launched the jets during the event. In the star-planet ILOT ZTF SLRN-2020 the star engulfed the planet rather than tidally disrupted the planet \citep{Deetal2023}. I suggested that the planet accreted some mass from the star in the months before the main outburst. At the early phase of the outburst as the planet spiralled-in close to the star the star tidally removed the accreted mass. This formed an accretion disk around the star that launched jets. This suggested process must be confirmed by three-dimensional hydrodynamical simulations. 

I also suggested that the planet launched jets as it accreted mass. These formed the slowly expanding outflowing dust. The jets that the planet launched form a concentrated outflow at $\approx 45 ^\circ$ to the equatorial plane. The jets that the star might have launched are perpendicular to the equatorial plane. Overall, I expect the outflow from ZTF SLRN-2020 to form a bipolar nebula, as other ILOTs have, e.g., Nova~1670 (CK~Vulpeculae; \citealt{Kaminskietal2020Nova1670, Kaminskietal2021Nova1670};  section \ref{sec:Jets}).    

\section*{Acknowledgments}

I thank Dima Shishkin and Amit Kashi for their help with the graphics.  I thank an anonymous referee for helpful comments. This research was supported by a grant from the Israel Science Foundation (769/20).

\section*{Data availability}
The data underlying this article will be shared on reasonable request to the corresponding author.  


\begin{thebibliography}

\bibitem[\protect\citeauthoryear{Bashi et al.}{2017}]{Bashietal2017} Bashi D., Helled R., Zucker S., Mordasini C., 2017, A\&A, 604, A83. 

\bibitem[\protect\citeauthoryear{Bear, Kashi, \& Soker}{2011}]{Bearetal2011} Bear E., Kashi A., Soker N., 2011, MNRAS, 416, 1965. 

\bibitem[Berger et al.(2009)]{Berger2009} Berger, E., Soderberg, A. M., Chevalier, R. A., et al. 2009, \apj, 699, 1850

\bibitem[Davidson, \& Humphreys(1997)]{DavidsonHumphreys1997} Davidson, K., \& Humphreys, R.~M.\ 1997, \araa, 35, 1

\bibitem[De et al.(2023)]{Deetal2023} De, K., MacLeod, M.,  Karambelkar, V. et al., 2023, Nature, 617, 55    

\bibitem[\protect\citeauthoryear{Gurevich, Bear, \& Soker}{2022}]{Gurevichetal2022} Gurevich O., Bear E., Soker N., 2022, MNRAS, 511, 1330. 

\bibitem[\protect\citeauthoryear{Herczeg et al.}{2016}]{Herczegetal2016} Herczeg G.~J., Dong S., Shappee B.~J., Chen P., Hillenbrand L.~A., Jose J., Kochanek C.~S., et al., 2016, ApJ, 831, 133. 

\bibitem[\protect\citeauthoryear{Holoien et al.}{2014}]{Holoienetal2014} Holoien T.~W.-S., Prieto J.~L., Stanek K.~Z., Kochanek C.~S., Shappee B.~J., Zhu Z., Sicilia-Aguilar A., et al., 2014, ApJL, 785, L35. 

\bibitem[Jencson et al.(2019)]{Jencsonetal2019} Jencson, J.~E., Kasliwal, M.~M., Adams, S.~M., et al.\ 2019, \apj, 886, 40

\bibitem[Kaminski et al.(2020)]{Kaminskietal2020Nova1670} Kaminski, T., Menten, K.~M., Tylenda, R., et al.\ 2020, arXiv:2006.10471

\bibitem[\protect\citeauthoryear{Kami{\'n}ski et al.}{2021}]{Kaminskietal2021Nova1670} Kami{\'n}ski T., Steffen W., Bujarrabal V., Tylenda R., Menten K.~M., Hajduk M., 2021, A\&A, 646, A1. 


\bibitem[Kaminski et al.(2018)]{Kaminskietal2018} Kaminski, T., Steffen, W., Tylenda, R.,  Young, K.~H., Patel, N.~A., \& Menten, K.~M.\ 2018, \aap, 617, A129


\bibitem[Kashi et al.(2019)]{Kashietal2019Galax} Kashi, A., Michaelis, A.~M., \& Feigin, L.\ 2019, Galaxies, 8, 2. 

\bibitem[Kashi \& Soker(2016)]{KashiSoker2016} Kashi, A., \& Soker, N.\ 2016, Research in Astronomy and Astrophysics, 16, 99

\bibitem[Kashi \& Soker(2017)]{KashiSoker2017Planet} Kashi, A. \& Soker, N.\ 2017, \mnras, 468, 4938. 

\bibitem[\protect\citeauthoryear{Livio}{2009}]{Livio2009} Livio M., 2009, ASSP, 13, 3 (Protostellar Jets in Context, by Kanaris Tsinganos, Tom Ray, Matthias Stute. Astrophysics and Space Science Proceedings Series. Berlin: Springer) 

\bibitem[\protect\citeauthoryear{Metzger, Giannios, \& Spiegel}{2012}]{Metzgeretal2012} Metzger B.~D., Giannios D., Spiegel D.~S., 2012, MNRAS, 425, 2778. 

\bibitem[Metzger, \& Pejcha(2017)]{MetzgerPejcha2017} Metzger, B.~D., \& Pejcha, O.\ 2017, \mnras, 471, 3200

\bibitem[\protect\citeauthoryear{Mobeen et al.}{2023}]{Mobeenetal2023} Mobeen M.~Z., Kami{\'n}ski T., Matter A., Wittkowski M., et al., 2023, 

\bibitem[Muthukrishna et al.(2019)]{MuthukrishnaetalM2019} Muthukrishna, D., Narayan, G., Mandel, K.~S., Biswas, R., \& Hlo{\v z}ek, R.\ 2019, \pasp, 131, 118002

\bibitem[\protect\citeauthoryear{O'Connor et al.}{2023}]{OConnoretal2023} O'Connor C.~E., Bildsten L., Cantiello M., Lai D., 2023, arXiv, arXiv:2304.09882. 

\bibitem[Pastorello \& Fraser(2019)]{PastorelloFraser2019} Pastorello, A., \& Fraser, M.\ 2019, Nature Astronomy, 3, 676
 
\bibitem[Pastorello et al.(2019)]{PastorelloMasonetal2019} Pastorello, A., Mason, E., Taubenberger, S., et al.\ 2019, \aap, 630, A75

\bibitem[Pejcha et al.(2016a)]{Pejchaetal2016a} Pejcha, O., Metzger, B.~D., \& Tomida, K.\ 2016a, \mnras, 455, 4351
  
\bibitem[Pejcha et al.(2016b)]{Pejchaetal2016b} Pejcha, O., Metzger, B.~D., \& Tomida, K.\ 2016b, \mnras, 461, 2527

\bibitem[Retter \& Marom(2003)]{RetterMarom2003} Retter, A., \& Marom, A.\ 2003, \mnras, 345, L25

\bibitem[Retter et al.(2006)]{Retteretal2006} Retter, A., Zhang, B., Siess, L., Levinson , A.\ 2006, \mnras, 370, 1573. 

\bibitem[Shara et al.(1985)]{Sharaetal1985} Shara, M.~M., Moffat, A.~F.~J., \& Webbink, R.~F.\ 1985, \apj, 294, 271

\bibitem[\protect\citeauthoryear{Sicilia-Aguilar et al.}{2017}]{SiciliaAguilaretal2017} Sicilia-Aguilar A., Oprandi A., Froebrich D., Fang M., Prieto J.~L., Stanek K., Scholz A., et al., 2017, A\&A, 607, A127. 

\bibitem[\protect\citeauthoryear{Soker}{2020}]{Soker2020JetsILOT} Soker N., 2020, ApJ, 893, 20. 

\bibitem[\protect\citeauthoryear{Soker \& Kaplan}{2021}]{SokerKaplan2021} Soker N., Kaplan N., 2021, RAA, 21, 090. 

\bibitem[Soker \& Tylenda(2003)]{SokerTylenda2003} Soker, N. \& Tylenda, R.\ 2003, \apjl, 582, L105. 

\bibitem[\protect\citeauthoryear{Yamazaki, Hayasaki, \& Loeb}{2017}]{Yamazakietal2017} Yamazaki R., Hayasaki K., Loeb A., 2017, MNRAS, 466, 1421. 

\end{thebibliography}
\end{document}